# Causal Stability Conditions for General Relativistic Spacetimes

E. M. Howard

*Department of Physics and Astronomy, Macquarie University, Sydney, NSW 2109, Australia*

**Abstract:** A brief overview of some open questions in general relativity with important consequences for causality theory is presented, aiming to a better understanding of the causal structure of the spacetime. Special attention is accorded to the problem of fundamental causal stability conditions. Several questions are raised and some of the potential consequences of recent results regarding the causality problem in general relativity are presented. A key question is whether causality violating regions are locally allowed. The new concept of almost stable causality is introduced; meanwhile, related conditions and criteria for the stability and almost stability of the causal structure are discussed.

## 1. Introduction

In the past several years, general relativity has been infiltrated with exotic geometries involving closed timelike curves [1] or other possible causal violations. For a given spacetime to be physically reasonable, the spacetime has to be robust against any possible perturbations of the metric. A good understanding of the nature of causality plays a fundamental role in the construction of any physical theory. The number of open issues concerning the causal behaviour of the spacetime is fairly large. As it is known, the causal properties of the spacetimes have been ordered in a hierarchy of conformally invariant features [2], a causal ladder of the spacetimes was established at the end of the 1960s through the works of Carter, Geroch [3], Woodhouse, Kronheimer, Seifert, Penrose and Hawking (Fig. 1). Roughly speaking, the author defines causality as "the relation between two events correlated in a regular pattern or between a cause an effect".

Any physical theory mutually assumes causation as an inherent fundamental assumption. In relativity theory, an event can influence another event only if there is a causal (timelike or null) geodesic curve connecting the two spacetime points.

Three different levels of comprehension of causal structure of the spacetime may be considered: the first one is an abstract-formal stage, with origins in special relativity, assigning a light cone to every single event in spacetime; the second stage has a topological nature and considers local differential behaviour of geodesics on a Lorentzian manifold; the third aspect assumes a global cosmological level of our understanding of causality and incorporates classical global problems in general relativity, as the initial value problem, spacetime boundaries or the singularity theorems.

## 2. Causal Stability

The stability of the spacetime properties has been matter of continuous interrogation since the birth of the exact solutions of Einstein equations. Stable causality is one of the most important global assumptions that

---

**Corresponding author:** Ecaterina Marion Howard, Research field: general relativity. E-mail: katie.howard@mq.edu.au.

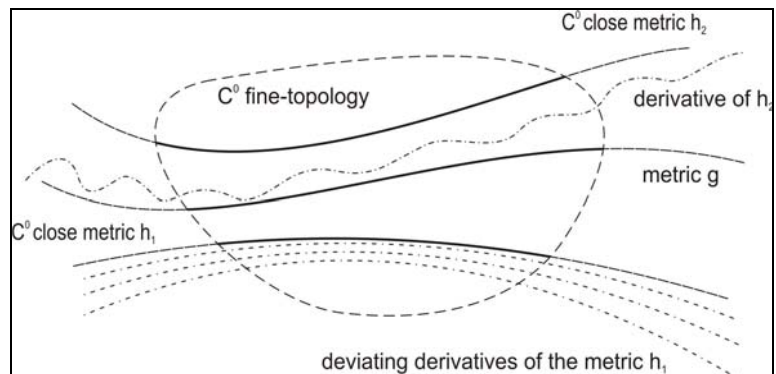

**Fig. 1  Metric perturbation in a $C^0$-fine neighbourhood.**

have been used in order to avoid dramatic alterations of the standard view in physics. The issue of causal stability is quite difficult to define as there are no given coordinate invariant methods able to define two metrics close enough to one another and consequently, there is no topology on a pseudo-Riemannian manifold that preserves coordinate invariance.

Causal stability is a primitive concept, more simple and basic than the notion of time function or chronology; it is assumed that it is a feature of the spacetime that can be derived entirely from properties of the topology of the spacetime and intimately related to the concept of closeness of metrics. The Whitney $C^r$-fine topology or just the $C^r$-fine topology on the space of metrics may be used, to provide a simple definition of the notion of $C^r$-stability or just causal stability.

Whitney $C^r$-fine topology is simply defined as a countably infinite family of topologies on the set of smooth mappings between two smooth manifolds. In other words, the Whitney $C^r$-topology gives us which subsets of a topology are open sets. For greater values of $r$, the topologies about a given metric, generated by a $C^r$ topology would become finer, as there are less metrics in any given neighbourhood. For example, in Fig. 1, the metric $h_2$ exists a $C^0$ topology, as its derivative is close to $g$, but it can not exist for higher values of $r$, because of a highly varying derivative. The first metric $h_1$ can exist in any $C^r$ as long as the neighbourhood chosen is small enough to include its derivatives. However, it looks unlikely, as its derivatives are not very close to $g$.

A spacetime $(M, g)$ is considered stably causal in defined on Whitney $C^0$-fine topology if there exists a $C^0$-fine neighbourhood, $Q_g$ of the Lorentzian metric $g$, such that for each $hQ_g$, $(M, h)$ is causal (Beem, Ehrlich and Easley). A definition of causal stability in a more primitive way can be now formulated, as the basic concept requires.

In this context, a stably causal spacetime will remain causal under small $C^0$-fine perturbations of the metric, in other words, stably causal spacetimes are causal for any metric, defined on a specific topology. Does causality require a "local" definition? Is causality a "local" concept that can not be defined on an entire manifold or as a generic global property of a specific topology?

In general theory of relativity, spacetime is considered as a generic collection of events which admits a 4-dimensional differentiable manifold structure associated with it. Recent research in spacetime singularities and boundary conditions has led to a strong development of causality theory and has brought to our attention various discussions around causal stability and physical reasonable conditions that could be imposed on the spacetime, in order to maintain it causally stable and robust, against any metric perturbations at infinity or in finite regions of the spacetime. A stably causal spacetime (as previously defined) still remains causal under small $C^0$-fine perturbations of any given metric.

However, if the spacetime is somewhat stable at a macroscopic level, what are the conditions that the spacetime structure should be confined to? Furthermore, how coarse should the spacetime topology be defined, in order to provide an accurate framework and set of conditions for causal stability.

Our main goal is to understand and investigate the main concept of causal stability and propose an alternative view and a more generalized picture of this fundamental property of the spacetime. Strong causality defined on a manifold does not necessarily ensure that the manifold does not contain causality violation regions [4]. If stable causality holds as well, the occurrence of causality violations is considerably diminished but still does not disappear. It is interesting here in announcing a fine edge between strong causality and stable causality and seek to fine tune stability by opening the light cones enough to affect the local properties of spacetime, including compactness but still keep the fundamental and global features, including causal stability of the entire manifold.

A new level in the hierarchy between strong and stable causality, which is called almost stable causality should be therefore weaker than stable causality and stronger than almost causality. Also, almost stable causality must have a purely topological origin.

The search for a topology associated with the new causal relation will constitute a topic of a different paper. Almost stable causality is defined in a trivial way as a local causal unstable and global causal stable feature of the spacetime. The main fundamental property of such a spacetime is the ability to open the light cones inside any subset of the spacetime, introducing locally closed causal curves but at infinity to "lose" any possible formed CTC and well behaving. Intuitively, the authors have a global hyperbolicity property, in initial Leray's formulation, not assigned to the whole spacetime, but only to any subset of the spacetime. Global hyperbolicity and asymptotic well behaviour becomes, somehow, a "local" property of the spacetime.

Can the authors open the light cones inside a "globally hyperbolic" subset of the spacetime or a compact set of causal curves, defined on a $C^0$-fine topology? The question is: what kind of topology would allow such a construct? Can Whitney $C^r$-topology help us here?

Two important causal conditions are pinpointed, related to popular criteria that have always been associated with the concept of "causal stability" and avoidance of classical physics paradoxes:

(a) Avoidance of "grandfather's paradox":

(1) Chronology: no closed timelike curves can exist;

(2) Causality: no closed causal curves can exist;

(3) Strong causality: no "almost closed" causal curves can exist (for each point and any neighbourhood of it, there exists a neighbourhood subset of the manifold, such that any causal curve with endpoints here is totally contained in the original manifold);

(4) Stable causality: close metrics to the original one are causal.

(b) Absence of "naked singularities" or avoidance of "information from one point to another escaping to infinity" (compactness of the diamonds).

In future work, all plausible and reasonable fundamental conditions that restrict the spacetime topological framework without spoiling causality will be analysed. A new level in the causal ladder that would eventually represent a central concept for understanding the interplay between local and global causal stability of the entire manifold will be introduced. Spacetime is considered stable if the light cones all over the spacetime without causing any instability or degeneracy on the spacetime can be opened, in other words without generating closed causal curves. Causality would be stable under small perturbations of the metric.

The key question here is: how much can authors relax this basic condition while preserving global causal stability and what are the consequences? Can the topology of the spacetime dictate or control how large these local perturbations of the metric can be, in order to avoid any disturbance or instability in any neighbourhood of our local changes?

How much will the spacetime allow us to vary a given metric while preserving global stability of the spacetime and at the same time provide a local reasonable stability as well?

How large can these perturbations be, to be able to preserve causality? Is there any topology that we can associate with such a new causality level in the hierarchy?

As an intuitive general relativistic picture, causality itself has been described so far by a number of trivial properties:

(1) The metric has to define global chronology or global ordering on the spacetime and admit a time function;

(2) The null cones of matter have to be included by the null cones of the metric, not allowing faster than light signals;

(3) The equations of motion have to admit a well-posed Cauchy problem.

It is interesting here in the maximal relaxation criteria of causal stability, which would represent a new important level in the causal ladder of the spacetimes. The level should be located between strong causality and stable causality and it will be called "almost stable causality".

Such criteria should be strong enough to assure global stability of the spacetime but locally would be sufficiently "relaxed" to generate instabilities on a small subset of the manifold, under a small perturbation of the metric.

The answer is almost stable causality, a new level is introduced in the causal hierarchy of spacetimes, as shown in Fig. 2. The new level would still allow causality violations at a macroscopic level.

Any perturbation of the metric should be small enough to hold continuity on the entire spacetime but high enough to allow discontinuity in an open neighbourhood at a local macroscopic level.

If the spacetime is almost causally stable, this peculiar property of spacetime would allow closed timelike curves, keeping the entire spacetime causally stable. It could be seen as a sub-property of causal stability, rather than a different standalone feature. But for reasons concerning causal hierarchy structure, this property will be added as a new separate level in the abstract framework.

This is a more "relaxed" model than the stably causal spacetime announced in 1974 by Hawking [5] as the most plausible global assumption to make.

Almost stable causality is the most fundamentally "restricted" plausible candidate on the macroscopic spacetime. Any further restriction would cause causal violations that would alter the global structure of the spacetime.

The authors shall try to provide an accurate definition of almost stable causality, as depicted in this paper. Firstly, recall the chronology condition. The condition holds on a spacetime $M$, if there are no closed timelike curves on $M$. In the same way, the causality condition holds on $M$ if $M$ does not admit closed causal curves.

A stronger condition, stable causality follows here. The stable causality condition holds on $(M, g)$, for a set of Lorentzian metrics $g$, if there is a neighborhood $Q_g$ of $g$, defined on Whitney $C^0$-fine topology, such that for any $h \cdot Q_g$, the chronology condition would hold on $(M, h)$.

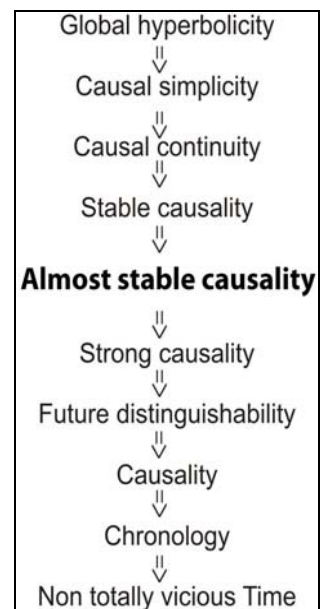

**Fig. 2   Causal hierarchy of spacetime-revised.**

A spacetime $M$ is almost stable causal if one of the following is true:

- There exists at least one open set $Q_g$ arbitrarily chosen, such that under small variations of the metric g, large enough to introduce closed causal curves on M, would not spoil stable causality outside $Q_g$ on the rest of $M$;
- The spacetime is globally causal stable and it is possible to widen the light cones on an open region of the spacetime, enough to introduce closed causal curves;
- The light cones can be open within the causal stable spacetime at a local and macroscopic level.

Almost stable causality property should be strictly topological and should be hidden in the fine topological properties of the spacetime, at both local and global level. In a physical sense, this new property of the spacetime should be defined as a continuous but not transitive function on a given subset and at the same time asymptotically decreasing with the distance, without spoiling causality on the rest of the manifold. In search for a reasonable topology that applies here, define almost stable causality outside a specific given topological framework, as Whitney fine topology was previously used. As it is known, a property of the spacetime can be stable in some topologies and unstable in others.

The property should be an asymptotic feature of the spacetime, defined in every point of the manifold and should be totally derived from a topological global property of the spacetime. Obviously, the "chronology condition" does not hold here, as the spacetime is not necessarily free from closed timelike curves. The presence of a cosmic time function [6] and a global time on such a spacetime will be subject of further work. What authors already know is that the stable causality property is a very strong condition, much too strong for a realistic spacetime, which led us to the new construct of almost stable causality. It is also known that a causally stable spacetime holds the chronology condition, therefore admits a time function. Does our construct necessarily need a time function? Is such a spacetime "realistic"? The questions still remain open.

## 3. Remarks

Is there a current rigorous definition of the causal structure of the spacetime? A generic acceptable definition should contain all the information about the Lorentzian manifold, causal properties and time orientability, a universal assumption always lying behind any physical theory. The main question about causality and its relation to time is here translated into an issue of assuring a perfect consistency of the causal ladder of spacetime.

The most important step in the causal hierarchy is the global hyperbolicity property, located at the top of the ladder. The concept is central to general relativity and generates several open questions in other context areas: initial value problem, singularity theorems, geodesic inextendability, imprisonment or causal boundaries. Geroch [3] has shown that if a spacetime is globally hyperbolic, it admits at least one submanifold that is intersected once by a an inextendible timelike geodesic.

Briefly, the spacetime is globally hyperbolic if it admits a Cauchy hypersurface. Any such a topological surface would be acausal if it is crossed only once by an inextendible timelike curve. In this sense, a lightlike geodesic is an inextendible achronal causal curve.

The standard definition of global hyperbolicity assumes two separate conditions:

(1) Strong causality (no "almost closed causal curves") [7] which was later weakened by Bernal, Sanchez (2000) [8] by replacing it with causality (no closed causal curves exist), by introducing a new concept of "causal simplicity";

(2) Compactness of the spacetime ("no naked singularities" condition), strictly derived from the

weak cosmic censorship conjecture.

Finally, it is worth pointing out that even though global hyperbolicity is a stable property (in the set of all time oriented Lorentzian metrics on a manifold), the causal structure of a globally hyperbolic spacetime can be unstable against local metric perturbations; it has been shown that the causal structure of Einstein and Minkowski static spacetimes remains stable under variations of the metric, whereas that of de Sitter becomes unstable. One of the other recent results is also statement that "chronological spacetimes without lightlike lines are stable causal" [9]. The physical meaning of this theorem is the fact that if the spacetime admits causality violations, then either chronology is violated or spacetime is singular. This means if almost stable causality is admitted as a fundamental distinctive topological property of spacetime, chronology is locally violated and it could be violated at a macroscopic level, without any global effect on spacetime.

If causal boundaries over the spacetime are considered, other possible open questions appear. Seifert [10] has proven that a globally hyperbolic spacetime always admits a causal boundary containing a timelike subset of causal curves. Again, a causal gradient is implicitly assumed in order to hold the consistency of the causal theory. If an unknown factor that defines some sort of continuous increasing function that applies to any future directed causal curve is applied, the consistency of the causal hierarchy of the spacetime is saved.

The implicit or explicit assumption of such a smooth time function containing a pre-ordering timelike gradient [11] has already been introduced by Hawking [12] in his work "the existence of cosmic time functions".

The cosmic time is defined as a global function that increases along every future directed timelike or null curve. The existence of such a function requires causal stability as a fundamental condition (no closed timelike or null curves in any Lorentz metric that is sufficiently near the spacetime metric).

Hawking [13] proves the equivalence between two fundamental features of the spacetime: stable causality and global time. Is it necessary to make all these tacit assumptions about a global time function, often associated with a cosmological flow? Can we speak about a global time? Are there any constraints in the local physical laws that would tell us anything about a global time?

The concept itself of a global time is taken for granted. The time function generates a total pre-ordering gradient on the spacetime manifold. It has been proven that under physically reasonable conditions, the absence of a global time function would imply a singular spacetime[14].

## 4. Conclusions

The main problem is that the authors do not have enough information about the evolution of the spacetime manifold from physically reasonable initial conditions and if this evolution would generate naked singularities.

There are no clear results that prove the existence of global hyperbolicity. If the spacetime is non-globally hyperbolic, the initial Cauchy data can not provide enough information from past time-like infinity to future infinity, to completely determine the current state of the universe.

If global hyperbolicity does not hold, information coming from spatial future null infinity should be taken into account. This concept could work if the authors modify the notion of causal precedence and disconnect it from the "hidden" timelike gradient self-contained in the causality definition.

From a deterministic point of view, the principle of causality states that a physical event described by various variables is fully determined at a given time by a previous event in the causal chain.

In the new definition, the cause still precedes the effect. However, this expression is separated from a "temporal" point of view. In this case, the cause would not necessarily precede the effect from past to future infinity. Old philosophical definitions of "final causality" could work here. Superluminal causal propagator in scalar quantum field would easily secure this concept.

A simple illustration of a non-causal theory is the elliptic Klein-Gordon scalar field, where information can propagate along a closed curve in the spacetime, any event along this curve being able to influence itself.

Another example could be the scalar field theory with a non-canonical kinetic term (or k-essence, MOND paradigm) and bi-metric MOND theories, allowing superluminal propagation of information, without threatening causality. The authors seek for a plausible causal property of the spacetime in which no prior chronology needs to be assumed. Almost stable causality is such a plausible candidate.

The authors basically want to stress here the fundamental role of causality without chronology (and orientability), as a fundamental property of the spacetime. The authors are looking for a generalized definition of causality that would not involve any chronology requirement in its expression (assuming a well posed Cauchy problem).

The assumption is a satisfactory mechanism that helps to climb the causal ladder, without violating Hawking's Chronology Protection Conjecture [15].

The "well behaved" Cauchy initial data would indeed contain information from spatial null infinity and it would lead to a "stable" state of the universe, without contradicting any regular physical assumptions (without violating either chronology or causality).

Whether the authors assume the existence of a global time function or a global hyperbolic spacetime, stable causality, as the strongest constraint in the causal ladder would be in this way saved.